\documentstyle[11pt,newpasp,twoside]{article}
\def\edcomment#1{\iffalse\marginpar{\raggedright\sl#1\/}\else\relax\fi}
\reversemarginpar

\begin{document}
\title{The dark side of star formation in galaxy clusters:
spectroscopic follow-up of clusters observed with ISOCAM}
\author{Dario Fadda} 
\affil{Instituto de Astrofisica de Canarias, Via
Lactea S/N, E-38200 La Laguna, Tenerife - Spain} 
\author{Pierre-Alain
Duc} 
\affil{CNRS (URA 2052), CEA Service d'astrophysique, F-91191
Gif-sur-Yvette Cedex, France}

\begin{abstract}
The evolution of galaxies in  cluster environments can be studied
using mid-IR observations which trace star forming regions hidden by
dust.  The optical follow-up of A1689 (z=0.2), observed at 6.7 and
15~$\mu$m by ISOCAM, have revealed a systematic excess of the B-[15 $\mu$m]
galaxy color distribution with respect that of Coma and Virgo
clusters. This result suggests the existence of a dark side of the
Butcher-Oemler effect measured in the optical. We present an analysis
of the optical/mid-IR properties of the mid-IR emitters in A1689,
comparing in particular the star formation rates based on mid-IR and
optical data. Morover, we present preliminary result for J1888, a
cluster at $z=0.56$ deeply observed with ISOCAM, based on recent
VLT/FORS and NTT/SOFI observations.
\end{abstract}

\section{Introduction}
Since the classical study of Butcher \& Oemler (1984) who put in
evidence an increasing fraction of blue, presumably star-forming,
galaxies as a function of cluster redshift, several authors have tried
to estimate the importance of the cluster environment on the evolution
of cluster galaxies.  Recent studies with sample of distant clusters
(Dressler et al. 1999, Balogh et al. 2000) pointed out that the star
formation rate (SFR) per cluster galaxy appears to be lower than that
in similar types of galaxies in the surrounding field.  The main
drawback of these studies is that the SFR rely on measurements of
optical line fluxes (usually [OII] or in a few cases H$\alpha$) which
suffer from strong dust extinction. Since dust obscuration depends on
object types and on environmental conditions, it is impossible to
correct it on average.  On the other hand, although galaxies affected
by extinction show typical spectral features (see Dressler et
al. 1999), it is quite impossible to quantify the total SFR on the
basis of the optical data alone.  Mid-IR surveys, along with radio
centimetric fluxes, provide the best estimates of the dust obscured
SFR (see e.g. Chary \& Elbaz 2001) when far-IR are not available. For
this reason a sample of 10 clusters at different redshift
($0.1<z<0.9$) has been observed with the ISOCAM camera on-board ISO.
We present here the optical/mid-IR properties of the galaxies detected
at 15 $\mu$m in the A1689 cluster (see Fadda et al. 2000 for ISOCAM
data) and preliminar results of the VLT follow-up program of J1888, a
cluster at $z=0.56$ deeply observed with ISOCAM.

\section{Hidden star formation in A1689}

\begin{figure}[t]
\plotfiddle{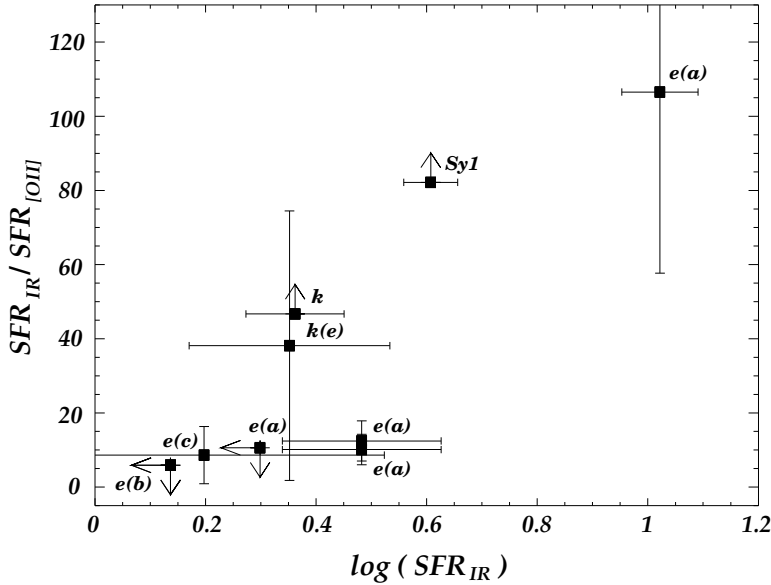}{7cm}{0}{40.}{40.}{-160}{0}
\caption{Ratio of the mid-IR to optical SFR estimates versus the
mid-IR based SFR estimate for A1689.  Note the high dust extinction in
some apparently passive galaxies ('k' type).}
\end{figure}

Using available images (HST and NTT) and about 100 spectra measured
with the NTT (Duc et al. 2001), we studied membership, morphologies
and spectral types of the objects detected by ISOCAM in A1689.  Our
analysis confirms the high fraction of blue galaxies in the cluster
( already detected by Butcher \& Oemler 1984) relying on a secure cluster
membership of the galaxies (spectroscopic and photometric redshifts
are known for all the galaxies observed).
Most of the 15 $\mu$m sources are luminous, blue, emission-line or
morphologically disturbed galaxies, i.e. a population of galaxies
usually associated with the ``Butcher--Oemler'' effect. However, $\sim$30\%
of the 15 $\mu$m sources do not show any sign of star-formation activity
in their optical spectrum.
More of 70\% of the emission-line galaxies in our spectroscopic sample 
are detected at 15 $\mu$m and all the galaxies classified as dusty starbursts 
( ``e(a)'' type in Dressler et al. 1999) are 15 $\mu$m sources. On the
contrary, none of the galaxies with a post-starburst optical spectrum have
been detected at 15 $\mu$m.

Since the AGN activity is very low in A1689, the 15 $\mu$m flux is a
reliable tracer of the dust-obscured star formation activity (see
Chary \& Elbaz 2001). Comparing the SFR based on the mid-IR (15 $\mu$m
flux) and optical ( [OII] line flux), we found that for galaxies with
mid-IR emission the ratio SFR(IR)/SFR(Opt) is very high and ranges
between 10 and 100, being the highest among ``e(a)'' galaxies (see
Figure~1). The median SFR(IR) is 2 $M_{\sun} yr^{-1}$, while the
median SFR(Opt) of the [OII] detected galaxies is 0.2 $M_{\sun}
yr^{-1}$. Although we are not observing luminous infrared galaxies
(the highest total IR luminosity $6.2\times10^{10} \ L_{\sun}$ is
measured in a ``e(a)'' galaxy), we conclude that a significant portion
of the star formation activity is visible only using mid-IR data.  At
least 90\% of the star formation is missed when estimated from the
[OII] line.  A comparison with the field, essential to investigate the
effects of the environment on the star formation of the cluster
galaxies, will be possible only when a statistical sample of mid-IR
galaxies for the coeval field of A1689 will be available.

\section{Preliminary results in J1888}

\begin{figure}
\plotfiddle{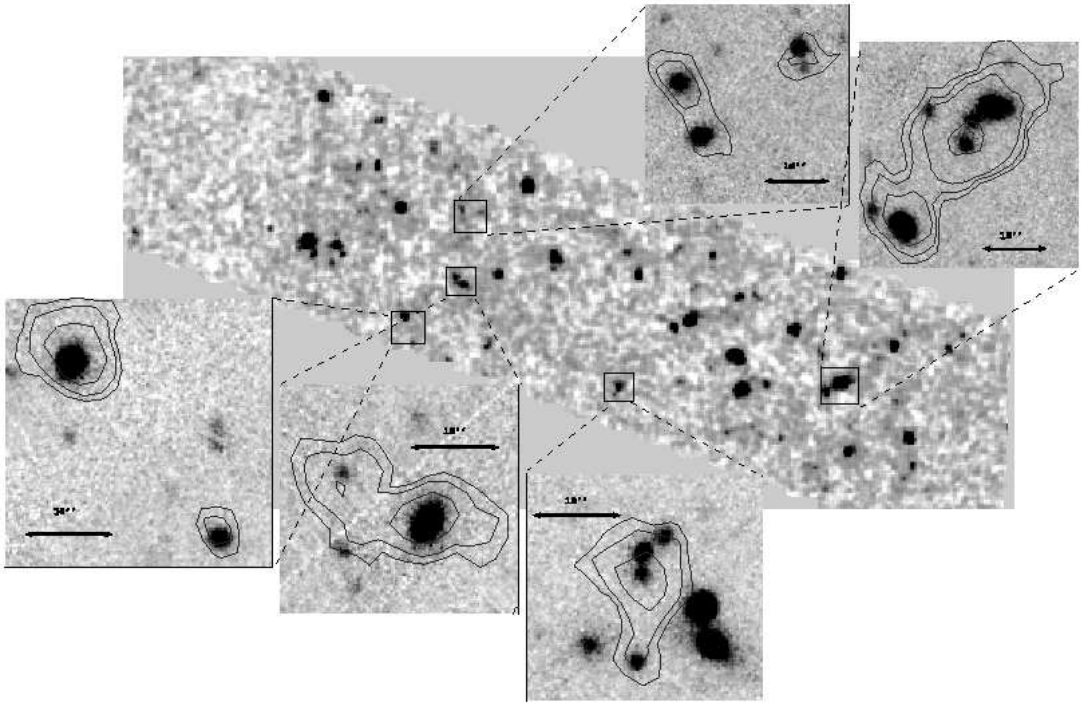}{10cm}{0}{40.}{40.}{-226}{-5}
\caption{15$\mu$m map of J1888 with more than 100 sources detected. Subsets
show 15$\mu$m overlays on optical images. Note the excellent astrometric precision of ISOCAM and the clear optical identifications.}
\end{figure}

J1888, a cluster at $z=0.56$, has been surveyed by ISOCAM at the same
depth of the Hubble Deep Field covering a strip of $14.5'\times3.5'$
crossing the cluster center in order to explore possible variation of
the mid-IR properties with the cluster-centric distance (see
Figure~2).  We have recently obtained about 120 spectra with VLT/FORS1
and a Ks image with NTT/SOFI, which complements B and R images obtained
with the 2.2m ESO telescope. So far we have redshifts for 55 out of 70
sources detected at 15 $\mu$m and 22 out of 47 sources detected at 6.7 $\mu$m.

The redshift distribution of the cluster, which appears very loose in the optical images, peaks at $z=0.56$ but the field is highly polluted by background galaxies. Considering the range of magnitudes of the cluster members (21.4 $< B <$25.3), 55\% of the galaxies do not belong to the cluster. Considering only the 15 $\mu$m sources, only 30\% of the ISOCAM sources are cluster members.

The color-magnitude relation is not obvious to determine even for
cluster members. Quite surprisingly there is a large number of ``red
outliers'' ($B-R >2$) which belong to the cluster and most interestingly several of them
are detected at 15 $\mu$m.  The presence of ``red outliers'' is not uncommon in clusters at $z\sim
0.5-0.6$ (see e.g. Margoniner \& de Carvalho 2000) but in this case
any contamination by field galaxies is ruled out.  Contrary to the case
of A1689, we have almost no mid-IR emitters among blue cluster members.

On the basis of VLT spectra we did a classification of galaxies
according to the scheme of Dressler et al. (1999). As in the case of
A1689, we detect at 15 $\mu$m all the dusty starbursts (``e(a)'' type)
and no post-starburst galaxies. Compared to A1689, we detect a larger
fraction of spiral-like spectra which is not unexpected for clusters
showing the ``Butcher-Oemler'' effect.

A detailed reanalysis of the J1888 ISOCAM data is now in progress using
the new method developed by Lari et al. (2001) in parallel with the analysis
of the deep ISOCAM surveys in the region of the Lockman Hole. The comparison
of the data from this survey, whose redshift distribution peaks at redshift
of 0.6 close to that of J1888, will be fundamental to study the influence
of cluster environment on the evolution of galaxies.

\end{document}